# Coded Path Protection
# Part 2: Design, Implementation, and Performance

Serhat Nazim Avci *Student Member, IEEE* and Ender Ayanoglu *Fellow, IEEE*

*Abstract*—In Part 1 of this paper, we introduced a coding-based proactive network protection scheme, named Coded Path Protection (CPP). In CPP, a backup stream of the primary data is encoded with other data streams, resulting in capacity savings. In addition to being a systematic approach of building valid coding structures, CPP is an optimal and simple capacity placement and coding group formation algorithm. It converts the sharing structure of any solution of a Shared Path Protection (SPP) technique into a coding structure with minimum extra capacity. In this Part 2 of the paper, we describe the implementation of our algorithm using Integer Linear Programming (ILP), its timing and synchronization requirements, and implementation issues in networks. We present simulation results which confirm that CPP provides faster link failure recovery than SPP while it incurs marginal extra capacity beyond that of SPP.

*Keywords*- Networks, network fault tolerance, codes, shared path protection, network coding.

## I. INTRODUCTION

In Part 1 of this paper, we introduced a novel proactive network protection scheme called Coded Path Protection (CPP). CPP is faster and more stable than rerouting-based schemes because it is proactive and it eliminates real-time configurations after a failure. The capacity placement algorithm of CPP is based on converting the sharing operation of Shared Path Protection (SPP) into encoding and decoding operations with an incremental extra cost. In this second part of the paper, Integer linear programming (ILP) is incorporated to carry out optimal conversion with minimum total capacity. CPP implementation requires synchronization of the data streams. We discuss this operation and, in addition, the effects of our algorithm and others for restoration time, stability, and signaling in networks. We performed comparisons between our scheme and SPP. To that end, simulations over realistic network scenarios using ILP formulations are carried out and their results are discussed.

## II. THE ILP FORMULATIONS

For a given network and traffic data, the primary paths are routed using the shortest paths. We developed an ILP formulation to find the optimal spare capacity placement (SCP) for the SPP technique under the wavelength continuity

The authors are with the Center for Pervasive Communications and Computing, Department of Electrical Engineering and Computer Science, University of California, Irvine, CA 92697-2625, USA.

This work was partially supported by the National Science Foundation under Grant No. 0917176. Any opinions, findings, and conclusions or recommendations expressed in this material are those of the authors and do not necessarily reflect the view of the National Science Foundation.

This work was presented in part during the IEEE International Conference on Communications, Ottawa, Canada, June 2012.

constraint. We also developed an ILP formulation to convert this SPP solution into a viable CPP solution.

As stated in [1], the problem of joint path routing and wavelength assignment is very complex. Therefore, for the purposes of this paper, the developed SCP solution of SPP is suboptimal. However, the ILP formulation of conversion from the SPP to the CPP algorithm is optimal. Note that the optimality of the conversion to CPP remains when an optimal SCP solution of SPP is employed, albeit finding that solution may be hard. The ILP formulation of the SPP inputs the primary paths found using shortest distances. The ILP formulation of SPP has the following set of parameters.

- $G(V, E)$: The network graph
- $S$: The set of spans in the network. Each span consists of two opposite directional links
- $N$: Enumerated list of bidirectional connections
- $T$: Maximum number of wavelengths on a link
- $c_e$: Cost (length) of span $e$
- $\Gamma_i(v)$ : The set of incoming links of each node $v$
- $\Gamma_o(v)$ : The set of outgoing links of each node $v$

There are two binary parameters, taken from the shortest path routing solution of the primary paths

- $x_e(i)$: Equals 1 iff the primary path of connection $i$ traverses over span $e$, is acquired from the shortest path routing
- $m(i,j)$: Equals 1 iff the primary path of connection $i$ is link-disjoint to the primary path of connection $j$, is acquired from the shortest path routing

There a number of binary variables

- $y_e(i)$: Equals 1 iff the protection path of connection $i$ traverses over span $e$
- $n(i,t)$: Equals 1 iff the protection path of connection $i$ uses the wavelength $t$ throughout the network
- $a_e(t)$: Equals 1 iff the wavelength $t$ on span $e$ is reserved by at least one protection path
- $k(i,j)$: Equals 1 iff the protection path of connection $i$ is link-disjoint to the protection path of connection $j$.

The objective function is

$$\min \sum_{t=1}^{T} \sum_{e \in S} c_e \times a_e(t). \qquad (1)$$

The origination, continuation and termination of protection paths are defined by

$$\sum_{e \in \Gamma_i(v)} y_e(i) - \sum_{e \in \Gamma_o(v)} y_e(i) = \begin{cases} -1 & \text{if } v = s_i, \\ 1 & \text{if } v = d_i, \\ 0 & \text{otherwise,} \end{cases} \qquad (2)$$

where $s_i$ and $d_i$ are the source and destination nodes of connection $i$, respectively.

We have

$$\sum_{t=1}^{T} n(i,t) = 1, \quad \forall i \in N. \qquad (3)$$

Equation (3) ensures that a protection path uses only a single wavelength to satisfy the wavelength continuity constraint.

In addition,

$$n(i,t) + y_e(i) \leq a_e(t) + 1, \\ \forall i \in N, \forall e \in S, 1 \leq t \leq T. \qquad (4)$$

In inequality (4), connection $i$ reserves the wavelength $t$ on span $e$ if the protection path of $i$ passes over the span $e$ and uses wavelength $t$ throughout the network.

Furthermore,

$$n(i,t) + n(j,t) \leq 1 + m(i,j) + k(i,j), \\ \forall i,j \in N, i < j, 1 \leq t \leq T, \qquad (5)$$

$$y_e(i) + y_e(j) + k(i,j) \leq 2, \quad \forall i,j \in N, i<j, \forall e \in S, \qquad (6)$$

$$y_e(i) + x_e(i) \leq 1, \quad \forall i \in N, \forall e \in S. \qquad (7)$$

Inequality (5) ensures that if two connections use the same wavelength on their protection path then their primary paths or protection paths must be link-disjoint. Inequality (6) checks if the protection paths of connection $i$ and connection $j$ are link-disjoint. Inequality (7) ensures that primary and protection paths of connection $i$ are link-disjoint.

In addition to the parameters of the SPP ILP formulation, the ILP formulation of the CPP has the following set of input parameters

- $C$: The maximum number of coding groups, depends on the average nodal degree of the network and size of $N$
- $y_e(i)$: Equals 1 iff the protection path of connection $i$ traverses over span $e$, is acquired from the solution of SPP

The parameter $C$ is similar to the parameter $T$ because the protection paths in the same coding group use the same wavelength throughout the network. The ILP formulation of CPP is very similar to the ILP formulation of SPP except for three differences. First, the parameter $T$ is replaced by parameter $C$. Second, $y_e(i)$ is taken as input from the SPP solution, and therefore, the origination, continuation, and termination of protection paths are not formulated again. Last, inequality (5) is replaced by

$$n(i,t) + n(j,t) \leq 1 + m(i,j), \\ \forall i,j \in N, i<j, 1 \leq t \leq C. \qquad (8)$$

Inequality (8) ensures that if two connections will be in the same coding group then their primary paths must be link-disjoint. Note that as a result, a single coding group can include multiple link-disjoint tree structures. These link-disjoint trees can be considered as individual coding groups. The ILP formulations are similar to that of [2] except the primary paths and protection paths are given as input depending on the scenario.

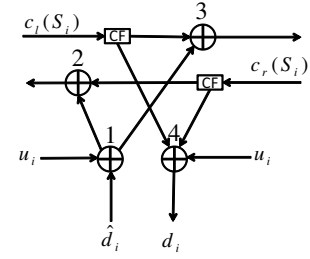

Fig. 1. Coding operations inside an end node.

## III. Synchronization and Buffering

In this section, implementation details about the encoding and decoding operations are given. In order to realize the coding structure over CPP topologies consisting of trees, synchronization and buffering are required. In [2], to help with the synchronization, the concept of round numbers are defined. The parity signals which are generated at the same time must be coded together at both ends of each link so that these signals can be decoded using the reciprocal signals from the opposite direction. Therefore, a round number is given to each parity data produced at a specific time instant. Round numbers keep track of the generation time of the parity signals. Due to propagation and transmission delays, the arrival times of parity signals, which will be coded together, are not necessarily the same at a node. Therefore parity signals must be synchronized using buffers at the nodes where encoding and decoding operations occur.

In contrast to [2], it is possible to employ round numbers as a signaling overhead only at the planning phase of the network. Once the parity signals are synchronized and the buffers are set up, then there is no need for a round number in front of each parity signal. If the network operation is disrupted somehow then round numbers can be employed to put the coding structure back on track.

### A. Synchronization

The goal of synchronization is to equalize the arrival time of the parity signals, which belong to the same round, to the end nodes. Encoding and decoding operations are carried out at the physical and hypothetical end nodes. Hypothetical end nodes physically refer to either a port of the real end nodes or an intermediate node over a linear coding trail. End nodes input three different signals in addition to the signal generated at this end node. Three encoding operations and one decoding operation occur in these end nodes. Since the same parity signals are used both for encoding and decoding, the end nodes adopt the "copy and forward" technique. In this technique, one copy of each parity signal is generated to use in decoding and the original copy is used for encoding purposes. Different copies of the same signal use different buffers. The detailed operations inside an end node $S_i$ is depicted in Fig. 1. In this figure, XOR 1, 2, and 3 are encoding operations and XOR 4 is the decoding operation. The box $CF$ copies and forwards the received data. For each XOR, the input signals must be synchronized by delaying the signals except the latest arriving signal.

## B. Buffering Delays

The input signals of XOR 1, $u_i$ and $\hat{d}_i$, are generated at the same time at $S_i$ and $T_i$, respectively. Therefore, in XOR 1, $u_i$ is delayed as much as the propagation delay of $\hat{d}_i$ over the primary path between $T_i$ and $S_i$. In XOR 2, it is assumed that the parity signals of each end node belonging to the same round are generated at the same time, which means signal $u_i \oplus \hat{d}_i$ arrives earlier than the other signal. The other signal $c_r(S_i)$ is defined as

$$c_r(S_i) = \bigoplus_{\substack{j=1 \\ S_j \in R}}^{N} u_j \oplus \hat{d}_j \ \oplus \ \bigoplus_{\substack{j=1 \\ T_j \in R}}^{N} \hat{u}_j \oplus d_j$$

where $R$ is the set of end nodes which remains at the right end side, if the tree topology is cut into two separate pieces at the end node $S_i$. The end nodes which remain at the left side compose the set $L$. The left and right side are chosen arbitrarily. There are two important properties with sets $R$ and $L$ that

$$R \cap L = \emptyset$$
$$R \cup L \cup S_i = \bigcup_{i=1}^{N} S_i \cup T_i.$$

Note that in $R$, some of the end nodes are coupled with other end nodes which belong to the same connection index, such as $S_j$ and $T_j$. The parity signals of these end nodes cancel each other with some extra delay. The upper bound of buffering delay of the signal $u_i \oplus \hat{d}_i$ at $XOR\ 2$ is

$$B_2 = \max_{S_p \in R \oplus T_p \in R} PD_p + \sum_{S_p \in R \wedge T_p \in R} CD_p.$$

$PD_p$ is the propagation delay from the end node $S_p$ or $T_p$ to the end node $S_i$. $CD_p$ is the canceling delay of signals which are coupled inside the set $R$.

In XOR 3, the buffering and synchronization are very similar to the one in XOR 2 except set $R$ is replaced by set $L$ and $c_r(S_i)$ is replaced by $c_l(S_i)$. The signal $c_l(S_i)$ comes later than $u_i \oplus \hat{d}_i$, which means the latter needs to be delayed by a buffer. The upper bound of buffering delay is very similar to the one in XOR 2 as

$$B_3 = \max_{S_p \in L \oplus T_p \in L} PD_p + \sum_{S_p \in L \wedge T_p \in L} CD_p.$$

XOR 4 handles the decoding operation where three different signals are input. Two of these signals are delayed by buffers in order to decode the signals belonging to the same round. Signal $u_i$ is available before the other two. If $B2 \leq B3$, then $c_r(S_i)$ and $u_i$ are buffered for $B3 - B2$ and $B3$, respectively. Otherwise, $c_l(S_i)$ and $u_i$ are buffered for $B2 - B3$ and $B2$, respectively.

## IV. RESTORATION TIME, STABILITY, AND SIGNALING

### A. Restoration Time

In this section, we conduct a qualitative and a quantitative analysis in terms of restoration time, stability, and signaling complexity of the SPP and the CPP approaches.

The developments in the optical XOR operations [3] allow coded path protection to be applicable in all-optical networks [4]. Wavelength assignment of CPP is trivial after converting the wavelength assignment solution of SPP because CPP is inherently suited to the wavelength continuity constraint. With this constraint, protection paths in the same coding group make use of the same wavelength throughout the network. SPP [5] is proposed for all-optical networks but some shared path protection techniques, such as [6], and the p-cycle techniques make use of "optical-electrical-optical" (o-e-o) conversion at intermediate nodes. The term all-optical or transparent optical networks have no o-e-o conversion at the intermediate nodes. The term opaque or translucent optical network is used for networks with optical transport that employ o-e-o conversion at some, but not all, intermediate nodes.

In CPP, the second (protection) copy of any data is generated and transmitted by the source node to the destination node after a fixed time delay. Protection in CPP is a proactive mechanism because only this data is used for failure recovery, without any feedback signaling. An advantage of proactive protection mechanism is the continuous operation over protection paths which means there is no need to configure and test an optical cross-connect (OXC) after any failure. OXC configuration and testing is the main source of delay in routing-based protection mechanisms [7]. In addition, this proactive mechanism eliminates the need of complex signaling and assures transmission integrity because the operations are all automatic. As stated in [7], transmission integrity can be the main problem in configuring protection paths and routing data over these paths in optical networks. This claim is supported by the stability concerns cited in [8].

Despite the fact that CPP is a proactive mechanism, it can utilize the signaling capabilities of opaque optical networks to make the recovery process faster. In some cases, a synchronization mechanism with different data streams can neutralize the time savings of CPP. For that purpose, we propose a two-tier protection mechanism available for opaque networks. Transparent networks need to stick with the proactive mechanism due to the weak signaling capability of all-optical networks. In the first step, protection is immediate as it is synchronized. As a second step, when the end nodes receive the error signals, they add one bit control message to the data signals, which transform them into "ambulance" signals. The "ambulance" signals skip the buffers and are encoded and decoded with data streams consisting of all zeros. The second step is similar to SPP but there is no need to configure the cross-connects. As a result, CPP is faster and more stable even in the worst case. The restoration time of the first part of the operation is

$$RT_{\text{CPP1}} = d_{sd} + h_b \times M + S,$$

where $d_{sd}$ is the propagation delay from end node $s$ to end node $d$ and $h_b$ is the number of hops in the protection path between $d$ and $s$. The symbol $M$ denotes the node processing time and varies on the type of optical network. It is taken as 0.3 ms in opaque networks [9] and 10 $\mu$s in transparent networks [5]. The symbol $S$ represents the delay due to synchronization.

It increases as the coding group size and the coding tree length increase. The restoration time formulation of the second step is

$$RT_{\text{CPP2}} = F + 2 \times d_{sd} + (h_{is} + 1) \times M + (h_b + 1) \times M,$$

where $F$ is the failure detection time and $h_{is}$ is number of nodes between node $i$, which detects the failure, and node $s$. The exact formulation of CPP for opaque optical networks is

$$RT_{\text{CPP}} = \min(RT_{\text{CPP1}}, RT_{\text{CPP2}}).$$

The recovery process in SPP starts with failure detection. Failure notification is required before end nodes switch the traffic from primary to protection paths. Intermediate nodes configure themselves after they receive error state signals. In the protection switching step, some researchers claim that nodes in the protection path configure the OXCs simultaneously which leads to significant restoration time savings [1]. Error state signals should be transmitted over a specialized control plane to notify every node to enable simultaneous configuration of cross-connects over the protection path. As a tradeoff, this incurs high signaling complexity throughout the network. The restoration time formulation of SPP is being debated, e.g., the results of some of the formulations [1, 5] do not match the numerical results in [10]. The OXC configuration time is stated to possibly be about 10 ms, but it is also reported to be as much as one second [8]. In addition, in [11] it is pointed out that an extra 40-80 ms is required only for signaling and reconfiguration, such as uploading maps. This means the OXC configuration time is not the only source of delay in SPP. Besides, the queueing delays in the restoration time formulation of SPP are ignored due to dependency of these values on the infrastructure and the application. Keeping the ambiguity in mind, we adopt the formula in [1] for the quantitative analysis of restoration time in SPP, assuming a separate packet-based control plane exists and it has the same topology with the network of interest. The symbol $X$ represents the OXC configuration and test time, so that

$$RT_{\text{SPP1}} = F + 2 \times d_{sd} + (h_{si} + 1) \times M + X + (h_b + 1) \times M.$$

If a specialized control plane does not exist, in other words, if in-band signaling is employed, then the OXCs cannot transmit the control message before they reconfigure themselves. This leads to higher restoration time due to the reconfiguration of OXCs in series. The formula for this case is adopted from [5]

$$\begin{aligned} RT_{\text{SPP2}} &= F + d_{sd} + (h_{is} + 1) \times M + (h_b + 1) \times X \\ &\quad + 2 \times d_{sd} + 2 \times (h_b + 1) \times M. \end{aligned}$$

*B. Stability*

Minimizing the number of cross-connects that need to be configured after an event of failure has been a subject of study for many researchers. The key point of the whole concept of the *p*-cycle depends on decreasing the number of real time cross-connect configurations to 2 from an arbitrary large number. It is considered to speed up the recovery process. In addition, in optical networks, pre-connection of the cross-connects before any failure state poses more advantages than only speed. In [7], the drawbacks of capacity sharing in all-optical mesh networks are investigated. Concerns are expressed for any type of optical networks, especially transparent or translucent optical networks. The translucent optical network is a compromise between transparent and opaque networks. Authors of [7] claim that transmission integrity cannot be guaranteed when arbitrary spare wavelength channels are connected into each other with real time on-the-fly configuration of the OXCs. Polarization, dispersion, power levels, amplifier-gain transients, and intermodulation are some of the sources of the defects that need to be taken care of while setting up a new connection in Dense Wavelength Division Multiplexing (DWDM) on-demand. It is known throughout the communications industry that all-optical networks have still a long way to satisfy full transmission integrity to a high capacity end-to end demand. Similar concerns are stated in [8] for the case of all-optical ultra long-haul networks. These networks have high capacity and the distances between nodes are very high so they are inclined to experience the defects of all-optical networks severely. In [8], it is argued that building a restoration connection out of nothing requires not only tuning the lasers and receivers and configuring the cross-connects but also requires triggering several feedback loop segments for power equalization. Activating restoration wavelengths after a failure changes the power profile over each span; therefore a complicated process needs to take place in order to stabilize the restoration process. In this paper, OXC configuration time is taken at most 10 ms [5], but in reality it can take up to seconds due to power management issues in the restoration path [8].

There are other studies that aim to minimize the real-time cross-connect configuration process. The concept of "pre-cross-connected-trails" (PXT) [6] is proposed to transform some of the sharing structure of the SPP scheme in order to pre-connect several cross-connects in the planning state of the network. This work was developed for opaque optical networks, but the same idea was later also applied to transparent optical networks in [11]. It minimizes but cannot zero the number of real time cross-connect configurations.

CPP solves the transmission integrity and stability problem fundamentally and proposes a faster protection than any of the mesh-based protection schemes except 1+1 and 1:1 protection. For a faster recovery technique, one can refer to diversity coding [12], which is also developed by the authors. In CPP, both primary and protection paths are fully connected and used for transmitting signals. In the primary paths, data packets are sent by themselves but their coded versions are sent over the protection paths. Each cross-connect is configured and tested, and the power profile of all the connections in the network is stabilized in the network planning phase. The real-time operations in the all-optical network after a failure can be eliminated completely, except some of destination nodes do protection switching to continue receiving data packets or temporarily terminate transmission to aid decoding. Protection switching is a much simpler operation compared to a cross-connect configuration and takes much less time. The introduction of "ambulance" signals and modifications over the link-disjointness criteria incurs some real-time operations and signaling inside the network. However, these operations do not endanger the stability and the integrity of the network.

## C. Signaling Complexity

It is known that transparent all-optical networks have disadvantages in signaling capabilities and performance monitoring. Therefore, the SPP technique requires a separate packet-based control plane to monitor the performance of the optical network and manage the restoration process through signaling [1]. The control plane is not all-optical, which is contrary to implementing an all-optical network. In the case of in-band signaling, cross-connects cannot be configured in parallel so the restoration time increases significantly. In both cases, signaling complexity is high to handle many dynamic operations throughout the network, such as uploading the routing maps. It also poses an overhead for the whole network operation. For SPP, recovery starts with the failure detection at the end nodes of the failed link. The control plane is responsible of detecting the failure. After detection, a failure indication signal is sent to the source nodes of the failed connections and they transmit setup signals over the pre-planned protection paths. The separate control plane allows transmitting setup signals over the protection path without configuring every cross-connect in the protection path. The complexity of this signaling operation depends on the number of connections traversing over the failed link. The fact that failure control signals and setup messages are processed in series at nodes increases the recovery time for some of the connections significantly.

One of the main advantages of CPP is the lack of the control plane in the basic mode of operation. The transparent optical network is sufficient to conduct the protection operations. First, there is no need to detect the failure in CPP because protection paths are employed at all times independent of the states of the primary paths. Second, setup signaling is only required in the planning phase of the network to synchronize the data signals and XOR operations. Once the protection is set up, there is no need for real-time signaling. Destination nodes of the affected connections do the failure detection and protection switching in an upper layer so it does not compromise the optical layer. Enabling protection paths to have common links with other primary paths in the same coding group does not incur any signaling because failure over the protection path can be derived via comparing the data from the primary path and from the protection path. In the context of opaque optical networks, recovery can be made faster by introducing one bit of error message as a prefix to each data packet that is affected by the failure. This introduces some signaling overhead but this overhead is very low compared to the rerouting-based techniques and their variations.

## V. SIMULATION RESULTS

In this section, we will present simulation results for link failure recovery techniques previously discussed, in terms of their spare capacity requirements and their worst case restoration time. The $p$-cycle protection is not involved in the simulations because it is not relevant in this comparison.

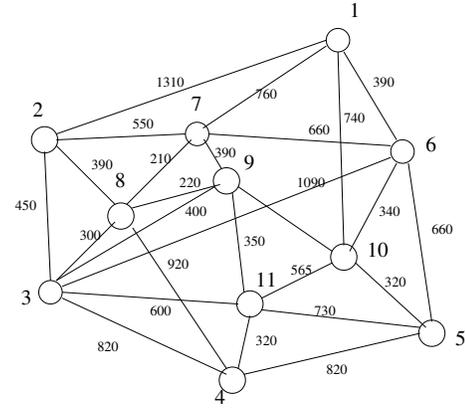

Fig. 2. European COST 239 network.

TABLE I
SIMULATION RESULTS OF COST 239 NETWORK FOR UNIFORM TRAFFIC SCENARIO

| COST 239 Network, 11 nodes, 26 spans | | | | | |
|---|---|---|---|---|---|
| Scheme | SCaP | RT for different X values (ms) | | | |
| | | 0.5ms | 1ms | 5ms | 10ms |
| CPP | 81.8% | 17.95 | 17.95 | 17.95 | 17.95 |
| SPP1 | 70.8% | 22.19 | 22.69 | 26.69 | 31.69 |
| SPP2 | 70.8% | 41.37 | 45.37 | 75.37 | 110.37 |

TABLE II
SIMULATION RESULTS OF COST 239 NETWORK FOR THE TRAFFIC SCENARIO OF [14]

| COST 239 Network, 11 nodes, 26 spans | | | | | |
|---|---|---|---|---|---|
| Scheme | SCaP | RT for different X values (ms) | | | |
| | | 0.5ms | 1ms | 5ms | 10ms |
| CPP | 85.2% | 14.89 | 14.89 | 14.89 | 14.89 |
| SPP1 | 75.0% | 19.93 | 20.43 | 24.43 | 29.43 |
| SPP2 | 75.0% | 37.23 | 39.73 | 59.73 | 84.73 |

The first network studied is the European COST 239 [13] network whose topology is given in Figure 2. In Fig. 2 and Fig. 3, the numbers associated with the nodes represent a node index, while the numbers associated with the edges correspond to the distance (cost) of the edge. The distances are useful to calculate the propagation delays. For the COST 239 network, there are two traffic scenarios, one uniform and the other, the scenario from [14]. SCaP represents spare capacity percentage

$$\text{SCaP} = \frac{Total\ Capacity - Shortest\ Working\ Capacity}{Shortest\ Working\ Capacity}$$

where *Shortest Working Capacity* is the total capacity when there is no traffic and the traffic is routed over shortest paths as explained in [15], and RT represents the worst-case restoration time. In order to cope with the high complexity of the path routing and wavelength assignment problem of SPP and to conduct a fair comparison between SPP and CPP, we adopted a three-fold strategy in simulations. In the first step, the demand matrix is partitioned into groups, which are as large as the memory limitations of the workstation allow. Second, the simple SPP algorithm without the wavelength continuity con-straint, taken from [10, p. 406], is run to obtain the protection paths of the connection in each partition. The primary paths of the connection demands are calculated using shortest-path routing and input to the simple SPP algorithm. In addition, the length limits for the protection paths of simple SPP solution is set to 4000 and 6000 kilometers in COST 239 and NSFNET networks, respectively. Third, we input the resulting primary and protection paths of the connection demands to the SPP algorithm with wavelength continuity constraint and the CPP algorithm, at the same time. Since partitioning demands cause randomness, we repeat the simulations 10 times and take their average. CPP results are obtained under the extended version of the link-disjointness criterion discussed earlier. We provide the SCaP values with the wavelength continuity constraint and restoration time results of both SPP and CPP in Table I and Table II for the two traffic scenarios, respectively. In these tables, SPP1 represents the case where there is a separate control plane (optical and electrical) while SPP2 is the case when in-band signaling (optical only) is used.

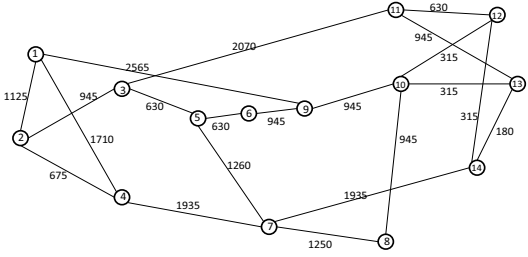

Fig. 3. NSFNET network.

TABLE III
SIMULATION RESULTS OF NSFNET NETWORK FOR UNIFORM TRAFFIC SCENARIO

| NSFNET Network, 14 nodes, 21 spans | | | | | |
|---|---|---|---|---|---|
| Scheme | SCaP | RT for different X values (ms) | | | |
| | | 0.5ms | 1ms | 5ms | 10ms |
| CPP | 119.1% | 37.15 | 37.15 | 37.15 | 37.15 |
| SPP1 | 107.1% | 49.18 | 49.68 | 53.68 | 58.68 |
| SPP2 | 107.1% | 80.93 | 83.93 | 107.93 | 137.93 |

TABLE IV
SIMULATION RESULTS OF NSFNET NETWORK FOR NONUNIFORM TRAFFIC SCENARIO

| NSFNET Network, 14 nodes, 21 spans | | | | | |
|---|---|---|---|---|---|
| Scheme | SCaP | RT for different X values (ms) | | | |
| | | 0.5ms | 1ms | 5ms | 10ms |
| CPP | 108.3% | 37.13 | 37.13 | 37.13 | 37.13 |
| SPP1 | 92.2% | 48.28 | 49.68 | 53.78 | 58.78 |
| SPP2 | 92.2% | 78.12 | 80.62 | 108.73 | 153.73 |

The second network studied is the NSFNET network [16], similar to the U.S. long-haul network [15]. Again, the traffic scenarios are uniform and nonuniform. In the nonuniform traffic scenario, there are 150 bidirectional connections which are generated using a gravity-based population model [17]. Same simulation procedure is applied for this network. We provide the SCaP values and restoration time results for SPP and CPP in Table III and Table IV.

As seen from the results, in each traffic and network scenario, the CPP solution results in 10-16% extra spare capacity percentage than the SPP solution. The results are consistent among different networks and traffic scenario meaning that capacity efficiency of CPP is competitive to that of SPP. On the other hand, the restoration speed increases three times over SPP2 when in-band signaling is used and increases approximately two times over SPP1 when there is a separate control plane for the SPP scheme. The restoration time of SPP increases as the expected time of OXC configuration and test increases. Realistically, in some cases it may take seconds. The time difference between these two techniques can increase if the reconfiguration, signaling, and queueing delays are accounted for. It is also observed that in the bigger region of the OXC configuration time, the restoration time of SPP2 is nearly six times larger than the restoration time of CPP.

The approach used in the simulations enables one to achieve close to optimal results in a short amount of time. There is still some potential to improve and achieve the optimal CPP results if the memory and the time limitations can be overcome. As a further research direction, the results of CPP can be improved by configuring the CPP algorithm to solve the routing problem of the protection paths itself without the input of the SPP algorithm.

## VI. CONCLUSION

In this paper, we introduced a proactive network restoration technique we call Coded Path Protection (CPP). The technique makes use of symmetric transmission over protection paths and link-disjointness among the connections in the same coding group. We modified the coding structure and leveraged its flexibility to convert sharing structure of a typical solution of SPP into a coding structure of CPP in a simple manner. With this approach, it is possible to quickly achieve optimal solutions. As a result of this operation, the CPP algorithm achieves significantly faster restoration. The restoration times for the competing algorithms (SPP1 and SPP2) vary with the OXC settling time, while they are independent of this parameter for our algorithm. As a result, the improvement in restoration time becomes more significant when the OXC settling time increases. Also, the improvement becomes substantial in the case of SPP2, or when in-band optical signaling is employed, or for all-optical networks. The penalty in spare capacity percentage is about 10-16%. With the availability of abundant fiber, this extra capacity does not constitute a significant burden on network resources. In addition, CPP has the advantages of full transmission integrity and stability and low signaling complexity.

In this paper, we have treated bidirectional links as such, as in Fig. 1(b) of Part 1 of this paper. An alternative is to treat bidirectional links as two unidirectional links, and develop a similar technique as in Fig. 1(a) of Part 1 of this paper. The timing requirements in such an approach are less stringent and therefore it is expected that the gains in restoration time will be even more significant. However, for this purpose, all of the SPP conversion theory of Part 1 of this paper needs to be revisited. We will conduct this study next.